\documentclass[aps,prb,amsmath,amssymb,
preprint,superscriptaddress]{revtex4-1}
\usepackage{hyperref}

\usepackage{graphicx}

\begin{document}








\title{Signatures of the Valley Kondo Effect in Si/SiGe Quantum Dots}










\author{Mingyun Yuan}
\altaffiliation[Present address: ] {Kavli Institute of NanoScience, Delft University of Technology, 2600 GA, Delft, The Netherlands}
\affiliation{Department of Physics and Astronomy, Dartmouth College, Hanover, New Hampshire 03755, USA}

\author{R. Joynt}
\affiliation{University of Wisconsin-Madison, Madison, Wisconsin 53706, USA}

\author{Zhen Yang}
\affiliation{Department of Physics and Astronomy, Dartmouth College, Hanover, New Hampshire 03755, USA}

\author{Chunyang Tang}
\affiliation{Department of Physics and Astronomy, Dartmouth College, Hanover, New Hampshire 03755, USA}

\author{D. E. Savage}
\affiliation{University of Wisconsin-Madison, Madison, Wisconsin 53706, USA}

\author{M. G. Lagally}
\affiliation{University of Wisconsin-Madison, Madison, Wisconsin 53706, USA}

\author{M. A. Eriksson}
\affiliation{University of Wisconsin-Madison, Madison, Wisconsin 53706, USA}

\author{A. J. Rimberg}
\email[]{ajrimberg@dartmouth.edu}
\affiliation{Department of Physics and Astronomy, Dartmouth College, Hanover, New Hampshire 03755, USA}










\date{\today}

\begin{abstract}

We report measurements consistent with the valley Kondo effect in Si/SiGe quantum dots, evidenced by peaks in the conductance versus source-drain voltage that show strong temperature dependence. The Kondo peaks show unusual behavior in a magnetic field that we interpret as arising from the valley degree of freedom. The interplay of valley and Zeeman splittings is suggested by the presence of side peaks, revealing a zero-field valley splitting between 0.28 to 0.34~meV. A zero-bias conductance peak for non-zero magnetic field, a phenomenon consistent with  valley non-conservation in tunneling, is observed in two samples.
\end{abstract}






\maketitle



The valley degree of freedom of conduction band electrons is one of several intriguing properties distinguishing silicon from III-V materials. The six-fold valley degeneracy in bulk Si is reduced to two-fold in Si/SiGe heterostructures due to the confinement of electrons in a two-dimensional electron gas (2DEG). The resulting valley splitting $\Delta$ in strained Si quantum dots (QD) and quantum point contacts is typically of order 0.2 meV.\cite{boykin2004,Goswami2007} A particularly interesting manifestation of valley physics would be the valley Kondo effect. The spin 1/2 Kondo effect comprehensively studied in GaAs quantum dots\cite{Goldhaber:1998,cronenwett1998,PhysRevLett.96.156802,Sasaki2000, PhysRevLett.88.126803, PhysRevB.67.113309,PhysRevB.72.165309} is usually observed when there is an odd number of electrons in the QD, in which the spin of an unpaired electron is screened by spins in the leads to form a singlet, resulting in a conductance resonance at zero dc bias.  When a magnetic field $B$ is applied, the QD spin states are split by the Zeeman energy $E_{z} = g \mu_{B}B$, $g$ being the Land\'{e} factor and $\mu_\text{B}$ the Bohr magneton.  The spin-Kondo resonance then splits into two peaks at $eV_\text{SD}=\pm E_{z}$, where $V_{\text{SD}}$ is the applied source-drain voltage.\cite{Goldhaber:1998,PhysRevLett.70.2601}   On the other hand, Kondo effects in Si/SiGe QDs have  only rarely been reported,\cite{Klein:2007} and there have been recent studies of dopants in a Si fin-type field effect transistor.\cite{PhysRevLett.108.046803} This is perhaps not surprising, since for the valley Kondo effect to occur, the energy associated with the Kondo temperature $T_\text{K}$ must be larger than the valley splitting $\Delta$, i.e. $k_\text{B}T_\text{K}>\Delta$ where $k_\text{B}$ is the Boltzmann constant, a rather stringent condition. Nonetheless, how the valley degeneracy in Si affects the Kondo effect in Si/SiGe QDs has been investigated theoretically\cite{PhysRevB.75.195345, PhysRevB.76.205314} and found to share some resemblance with carbon nanotubes.\cite{Herrero2005} The addition of the valley degree of freedom allows for a new set of phenomena to emerge, since both spin and valley indices can be screened. 
Measurement of the valley Kondo effect could therefore help probe the nature of valley physics in Si/SiGe QDs, a topic of great importance considering their potential application in quantum computation.\cite{Zwanenburg:2013} In particular, the question of how the valley index of an electron changes\cite{gamble} as it tunnels on and off a QD can be illuminated.\\

Here we report measurement of unconventional Kondo effects in two different Si/SiGe QD samples, one in perpendicular magnetic field $B_\perp$ and both in in-plane magnetic field $B_\parallel$. In both samples we observe conductance enhancement in two consecutive Coulomb diamonds, forbidden in a purely spin 1/2 Kondo effect. The resonances disappear as the temperature is raised above a few Kelvin. For Sample~I, in one diamond there are two peaks at zero magnetic field which coalesce and broaden into a single peak as $B_\perp$ is increased. In an adjacent diamond, a zero-bias resonance persists for finite $B_\perp$. For a different dot configuration of Sample~I measured in $B_\parallel$, we observe similar persistent zero-bias resonances in two consecutive diamonds. For Sample~II, there is a single resonance in one diamond which again does not split in non-zero $B_\parallel$. In an adjacent diamond we observe additional side peaks that converge and part again as $B_\parallel$ is increased. We argue that these counterintuitive phenomena may arise from an interplay between the valley and spin degrees of freedom.\\

\begin{figure}

\includegraphics{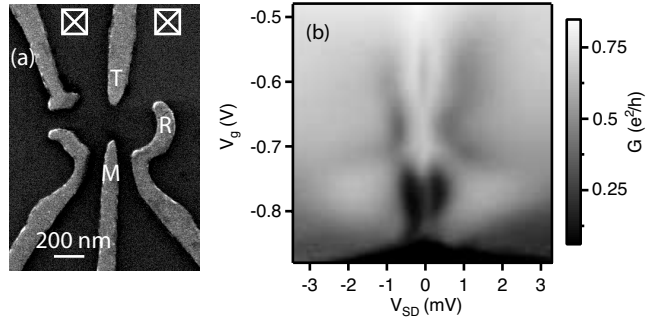}

 \caption{\label{fig1}(a) Electron micrograph of a Si/SiGe QD with identical design to the measured device. Crossed squares represent the ohmic contacts. (b) Stability diagram of differential conductance for  $V_M=-0.16$~V and $V_R=-0.53$~V  vs. bias voltage $V_{\text{SD}}$ and gate voltage $V_{\text{g}}$ applied to gate T at $B_\perp=0$.}

\end{figure}

The Si/SiGe heterostructure is grown using chemical vapor deposition (CVD). A step-graded virtual substrate is grown on Si (001) that was miscut 2~degrees towards (010).  A 1~$\mu$m thick Si$_{0.7}$Ge$_{0.3}$ buffer layer is deposited next, followed by an 18~nm Si well where the two-dimensional electron gas (2DEG) is located. A 22 nm intrinsic layer, a 1 nm doped layer ($\sim10^{-19}$ cm$^{-3}$ phosphorous), a second intrinsic alloy layer of $\sim$ 50 to 75~nm, and last a 9~nm Si cap layer are grown subsequently. The devices are fabricated with low-leakage Pd gates.\cite{yuan2011} Both QDs have the same design as shown in Fig.~\ref{fig1}(a), formed by gates T, M, and R, and their differential conductance is measured with lock-in techniques. We perform measurements in two dilution refrigerators. In one refrigerator the sample is oriented perpendicular to the magnetic field while in the other it is aligned in-plane with the field. The base temperature corresponds to an electron temperature of $T_e\approx150$ mK in both refrigerators.\\

We begin by discussing measurements of Sample~I in perpendicular magnetic field $B_{\perp}$ using an ac excitation of 3~$\mu$V. Gates M and R in Fig.~\ref{fig1}(a) are kept at $V_M=-0.16$~V and $V_R=-0.53$~V, respectively, while the voltage $V_g$ applied on gate T is varied. Interesting conductance enhancements appear in the Coulomb blockade region, as can be seen in the stability diagram of the QD differential conductance $G$ versus the dc bias voltage $V_\text{SD}$ and the gate voltage $V_g$ shown in Fig.~\ref{fig1}(b). There is a lower Coulomb diamond between $V_g\approx-0.85$ and $-0.69$~V and an upper one starting at $V_g\approx-0.69$~V and extending towards the top of the figure. The relatively large size of this lower diamond ($V_\text{SD}$ between $\pm3$ mV) suggests that the QD is in the few-electron regime, as observed in other Si/SiGe QDs of similar lithographic size and gate layout.\cite{Thalakulam:2010,Shi:2013}   The lower diamond also shows inelastic co-tunneling features consisting of two roughly vertical regions of enhanced conductance for $V_\text{SD}< -0.7$ mV and $V_\text{SD}>0.7$ mV that depend only weakly on $V_g$. The co-tunneling is due to virtual transitions to an excited state with excitation energy of about 0.7 meV, indicating that it involves a change in orbital state rather than valley state, for which the excited state energies are typically smaller than $\sim0.3$ meV. We make use of the co-tunneling feature's implications for electron number $N$ in the QD \cite{PhysRevLett.86.878} to estimate the charge states of the Coulomb diamond. Each energy level can be doubly occupied. For $N$ odd, elastic co-tunneling can always happen via an energetically favorable singly-occupied level instead of a higher level. Therefore, there will not be an abrupt enhancement of conductance at a certain bias voltage in the Coulomb blockade. For $N$ even on the other hand, for a co-tunneling event to take place  an electron must virtually occupy a higher energy level, which can only happen when the bias voltage reaches the threshold determined by the energy level spacing.  The results is a sudden step in the conductance at a non-zero bias. Thus the presence of inelastic co-tunneling in the lower diamond identifies it as corresponding to an even-electron-number state. The upper diamond must therefore correspond to an odd-electron-number state. We refer to the upper (lower) diamond as the odd (even)-number diamond. The Kondo effect in Si/SiGe QDs is forbidden for $N=4m, m=0,1,2,...,$ since the spin and valley states for a particular orbital state are filled.\cite{PhysRevB.76.205314}  Therefore, the even-number diamond corresponds to $N=4m+2$ while the odd-number diamond corresponds to $N=4m+3$.\\

\begin{figure}

\includegraphics{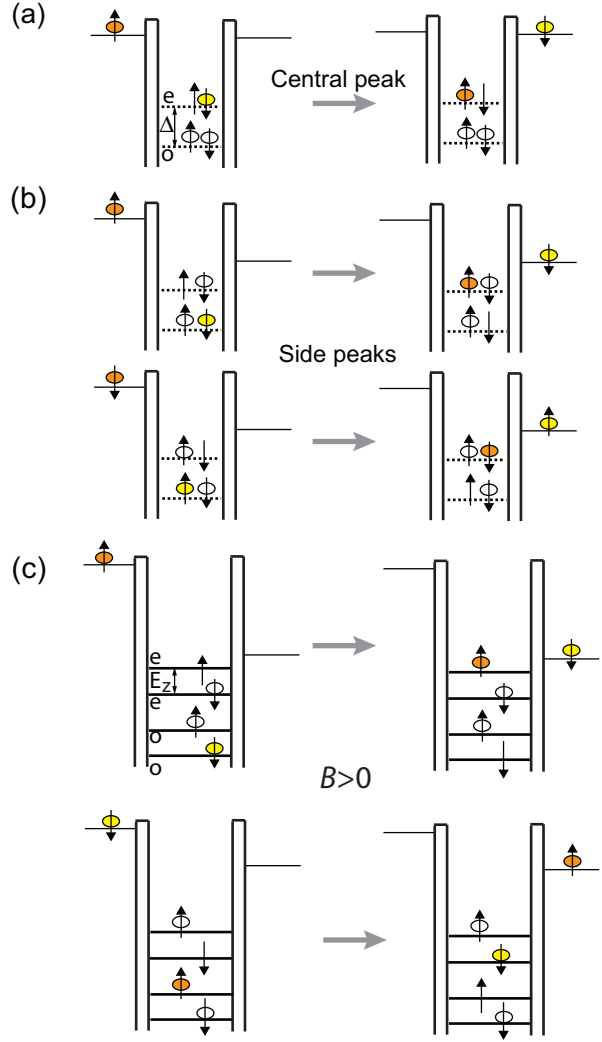}

 \caption{\label{fig3} Single-particle processes that can be associated with each conductance peak in the odd-number diamond. The many-body wavefunction is a combination of terms each of which is also associated with a particular process. (a) This process flips spin only and would give rise to a peak at zero bias, which is not observed. (b) The side peaks come from processes such as these that involve valley exchange. That the center peak is missing and the side peaks are present  implies $V_{ikm(\bar{m}=e)\sigma} < V_{ikm(\bar{m}=o)\sigma}$. (c) The same valley-reversing process in non-zero magnetic field, showing that each side peak is further split, explaining the broadening and eventual disappearance of these peaks as the field is increased.}
 \end{figure}

The Hamiltonian of a Kondo system with valley and spin degrees of freedom can be expressed as
\begin{equation}\label{Hval}
\begin{split}
H&=\sum_{ikm\sigma}\epsilon_{km\sigma}c^\dagger_{ikm\sigma}c_{ikm\sigma}+\sum_{m\sigma} \epsilon_{m\sigma} a^\dagger_{m\sigma} a_{m\sigma}+U\sum_{m'\sigma'\neq m\sigma}n_{m'\sigma'} n_{m\sigma}\\
&+\sum_{ikm\bar{m}\sigma}(V_{ikm\bar{m}\sigma}c^\dagger_{ikm\sigma}a_{\bar{m}\sigma}+\text{H.c.}),
\end{split}
\end{equation}
where $c^\dagger_{ikm\sigma}$ $(c_{ikm\sigma})$ creates (destroys) an electron with momentum $k$, valley index $m\in o, e$, spin $\sigma\in\uparrow,\downarrow$ and energy $\epsilon_{km\sigma}$ in lead $i\in L, R$, and $a^\dagger_{m\sigma}$ $(a_{m\sigma})$ creates (destroys) an electron with valley index $m$, spin $\sigma$ and energy $\epsilon_{m\sigma}$ on the QD\@. The first two terms in (\ref{Hval}) describe the energy of electrons in the leads and the QD, respectively.  The third term is the Coulomb interaction, which is taken to forbid double occupancy, and the fourth term describes the transfer of electrons between states in the the leads with valley index $m$ and in the QD with valley index $\bar{m}$. When $V_{ ik(m=\bar{m})\sigma}<V_{ik(m\neq\bar{m})\sigma}$, the valley mixing is relatively strong. The ground state of $H$ is a combination of terms that can be visualized graphically as processes that involve the hopping of an electron from the left lead to the right lead. Each distinct process can give rise to a peak in the conductance if the amplitude of the process is large enough while unobserved processes indicate a small tunneling matrix element. The position of the peak can be read off from the energy change of the electron in the process, while the height of the peak depends on the tunneling amplitude that produces it.  The processes resulting in the Kondo effect for a QD with $N=4m+3$ are illustrated in Fig.~\ref{fig3}, assuming that the valley splitting $\Delta$ varies slowly with magnetic field compared to $E_{z}$. Theory predicts three peaks for $B=0$ when the valley index is conserved: a peak at zero bias that involves a spin flip, commonly observed in GaAs QDs, and two valley side peaks that are due to changes in both spin and valley indices as the electrons tunnel through the QD.\cite{PhysRevB.75.195345} A magnetic field should split each valley Kondo peak into three peaks, the distance between neighboring peaks being $E_{z}/e$. In Fig.~\ref{fig3} two valley states with energy difference $\Delta$ are labeled by $o$ and $e$ (the choice of $o$ as the lower state is arbitrary). Note, however, that since parity is not a good quantum number in the device, the states are not necessarily odd and even in the z-coordinate.\\

\begin{figure}

\includegraphics{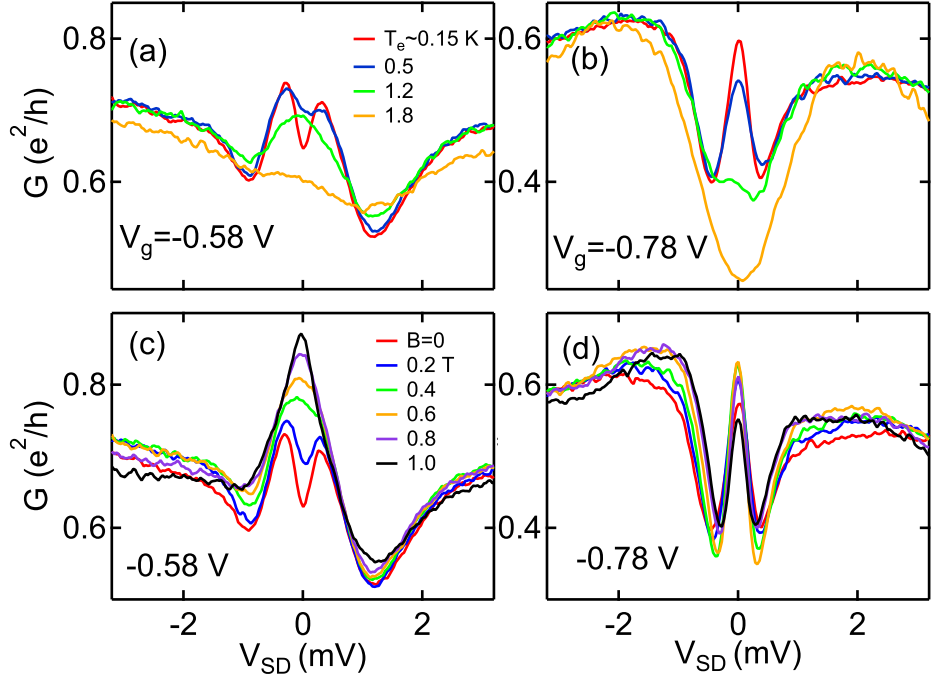}
 \caption{\label{fig2}(a),(b) Zero-field conductance $G$ vs.\ $V_\text{SD}$ in the odd(even)-number diamond taken at different temperatures. (c),(d): Magnetic field dependence of the Kondo peaks in (a),(b) at $T_e=0.15$~K.}

\end{figure}

The temperature dependence of the resonances in the odd- and even-number diamonds is shown in Fig.~\ref{fig2}(a) and (b), respectively, while their dependence on perpendicular magnetic field $B_{\perp}$ is shown in Fig.~\ref{fig2}(c) and (d). As can be seen in Fig.~\ref{fig2}(a) and (b), the resonances weaken as the temperature increases and vanish for $T>1.8$ K, verifying the presence of Kondo physics. There are several notable aspects of these resonances that contradict the spin-1/2 Kondo picture, such as a split resonance that coalesces into a single peak in the odd-number diamond and a single resonance in the even-number diamond that persists in $V_\text{SD}$ as $B_\perp$ is increased. These seemingly counterintuitive phenomena are consistent with valley Kondo physics in the QD.\\

In the odd-number diamond, we find two conductance peaks, one on each side of zero dc bias, suggesting the presence of the valley effect, as illustrated in Fig.~\ref{fig3}(b). Their position with respect to $V_\text{SD}$ stays mostly the same within the diamond. We fix $V_g=-0.58$V, varying the temperature (Fig.~\ref{fig2}(a)) and magnetic field (Fig.~\ref{fig2}(c)). At $T_e\approx150$ mK, the two peaks are clear. As temperature is raised, the peaks broaden and their height decreases. At 1.2 K only a single broadened peak can be resolved; it eventually becomes difficult to discern at 1.8 K. Two characteristic temperatures are roughly estimated from the data in Fig.~\ref{fig2}(a): $T_\text{K1}\approx2$ K for the Kondo feature to appear and $T_\text{K2}\approx1$ K to resolve the individual valley Kondo peaks.\\

Turning to the field dependence of these features in Fig.~\ref{fig2}(c), as $B_\perp$ is increased the two peaks broaden and coalesce, becoming indistinguishable at $B_\perp$=1 T.  Associated tunneling processes are illustrated in Fig.~\ref{fig3}(c). Note that the $g$ factor of Si is approximately 5 times larger than that of GaAs so that 1~T in these measurements is comparable to 5~T in GaAs QD experiments. The corresponding size of the Kondo peak splitting\cite{PhysRevLett.70.2601} is approximately $2E_z/e=2g\mu_B B/e\approx 0.23$~mV.  \\

In Fig.~\ref{fig3}(a) ($B=0$), the conventional spin 1/2 Kondo effect that results in a central peak is illustrated; note that the processes involve tunneling to and from only an $e$ valley state. In contrast, Fig.~\ref{fig3}(b) shows a process in which an electron occupying the $o$ valley state tunnels off the QD, and another electron from the lead tunnels on to the QD to occupy the $e$ state, resulting in the side peaks. The absence of a central peak in our experimental results suggests that the tunneling involving an odd valley is stronger than the pure even-valley tunneling and the Kondo temperature corresponding to the inter-valley process is higher than the spin-1/2 process. Therefore, at finite temperature, the side peaks are easier to observe while the center peak is obscured.\\

In this picture, the separation in $V_\text{SD}$ of the two side peaks is twice the zero-field valley splitting $\Delta/e$ in the QD, so $\Delta\approx0.28$ meV. A magnetic field lifts the spin-degeneracy of each valley state. In Fig.~\ref{fig3}(c) ($B>0$), each valley state splits into two spin states with energy difference $E_z$. A spin-down (up) electron in the $o$ state tunnels off the QD while a spin-up (down) electron tunnels on to the $e$ state, swapping both valley and spin indices. The other two processes remain degenerate, swapping only the valley index while preserving the spin index. As a result, the four processes generate three resonances in finite $B$ field. The resulting splitting broadens the two Kondo peaks and they coalesce into one peak, as shown in Fig.~\ref{fig2}(c).\\

\begin{figure}
\includegraphics{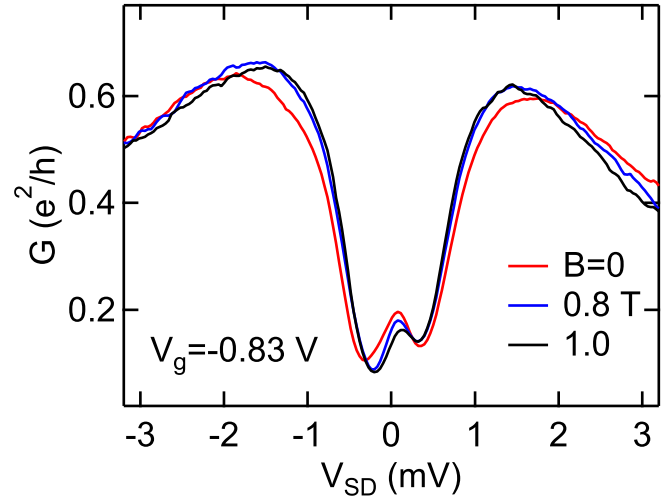}

\caption{\label{S1vsB} Kondo resonance for Sample~I  at $V_g=-0.83$~V in Fig.~\ref{fig1}(b) for three different values of perpendicular magnetic field.}

\end{figure}

In the even-number diamond, there is a single resonance at zero bias. As shown in Fig.~\ref{fig2}(b), the height of the resonance decreases monotonically with temperature, vanishing at $\sim$1.8 K. Surprisingly, as $B$ is increased this central resonance neither shifts nor splits, as shown in Fig.~\ref{fig2}(d), which is significantly different from the Kondo effects previously reported in semiconductor QDs. To determine if a peak splits in non-zero magnetic field, we  compare its full width at half maximum (FWHM) at zero field with the expected Kondo peak splitting $2E_z/e$. A peak is determined to be persisting near zero bias if double-peak structure is not observed at $2E_z/e\sim\text{FWHM}$. For example Fig.~\ref{S1vsB} shows the magnetic field dependence of the Kondo resonance corresponding to $V_g=-0.83$~V in Fig.~1(b). At 1~T the Kondo peak splitting corresponds to approximately  $2E_{z}/e\approx 0.23$~mV for $g=2$ and the FWHM of the black curve in Fig.~\ref{S1vsB} is about 0.24~mV, so we determine that it does not split. In fact, a peak at zero dc bias that persists in a non-zero magnetic field is expected to be a signature of a pure valley Kondo state and is associated with processes in which the valley index is not conserved during tunneling.\cite{PhysRevB.75.195345, PhysRevB.76.205314} For $B_\perp\neq0$ the center peak involving only valley screening dominates while the side peaks are suppressed. This suggests that the Kondo temperature relating to the pure valley process is greater than that of the spin process, a result of strong valley mixing $V_{ ik(m=\bar{m})\sigma}<V_{ik(m\neq\bar{m})\sigma}$. Also, with the magnetic field perpendicular to the 2DEG, we expect some dependence of the tunneling matrix elements on $B_\perp$, which might also be the reason why the peak height as well as the co-tunneling features show non-monotonic dependence on $B_\perp$. The data taken in $B_\parallel$ discussed in the following paragraph show monotonic field dependence.\\

\begin{figure}

\includegraphics{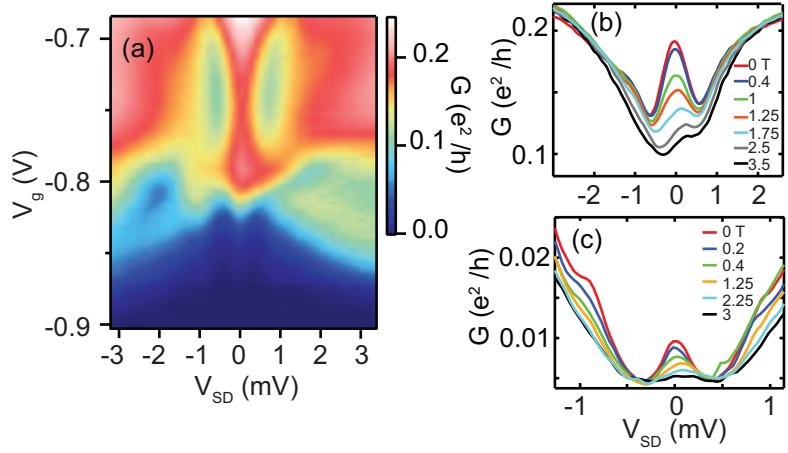}

 \caption{\label{S1}(a) Stability diagram showing the Kondo effect in two adjacent Coulomb diamonds for Sample I in $B_\parallel$ and in a different dot configuration.  Here,  $V_M=-0.6$~V and $V_R=-0.02$~V, while $V_{g}$ is applied to gate T\@.  (b) and (c) $B_\parallel$ dependence of the Kondo resonances for $V_g=-0.75$~V and $-0.83$ V in (a), respectively.}

\end{figure}

Sample I is also measured in a parallel magnetic field and once again we observe non-splitting Kondo peaks, this time for a different dot configuration. By measuring the QD in $B_\parallel$ instead of $B_\perp$ the dependence on magnetic field of the tunnel matrix elements is reduced. Fig.~\ref{S1}(a) shows the stability plot of two consecutive charge states formed  in this second dot configuration for $B_\parallel = 0$. Gates M and R are kept at $V_M=-0.6$~V and $V_R=-0.02$~V respectively while gate T is swept by $V_g$. A Kondo resonance at zero dc bias emerges in both charge states. When $B_\parallel$ is increased, the center peaks in both diamonds persist without splitting even for $B\geq3$~T, as demonstrated in Fig.~\ref{S1}(b) for $V_g=-0.75$~V and (c) for $V_g=-0.83$~V. Since the $g$ factor in Si is about 5 times bigger than that in GaAs, 3~T here is comparable to 15~T for a GaAs based QD. The height of the peaks decreases monotonically with $B_\parallel$.\\

The appearance and persistence of the center peak with $B>0$ is a key prediction of the theory of the valley Kondo effect and arises physically from an interference of valley conserving and valley non-conserving processes. This suggests that valley index is not always conserved during tunneling, a subject of some debate,\cite{Herrero2005,PhysRevB.75.195345,PhysRevLett.108.046803,PhysRevB.81.115324,PhysRevB.82.205315,Friesen:2006,Saraiva:2011,JiangZ:2012,JiangY:2013} and in accord with recent effective mass calculations of the effects of atomic step disorder at the quantum well interface.\cite{gamble} Secondly, given a non-zero $\Delta$, the QD electrons must  occupy an excited state rather than the ground state viewed in terms of single-particle levels. This is allowed energetically since by forming the many-body state with the leads, the system gains energy.  From the FWHM of the central peak in Fig.~\ref{fig2}(d) we estimate this energy gain is on the order of $k_\text{B}T_\text{K}\approx0.4$ meV, in good agreement with the temperature at which the resonance disappears. Assuming $\Delta\approx0.28$~meV in the even-number diamond in Fig.~\ref{fig1}(b), similar to that in the odd-number diamond, we have $\Delta<k_\text{B}T_\text{K}$, recovering the condition for the valley Kondo effect to be observed.\\

\begin{figure}
\includegraphics{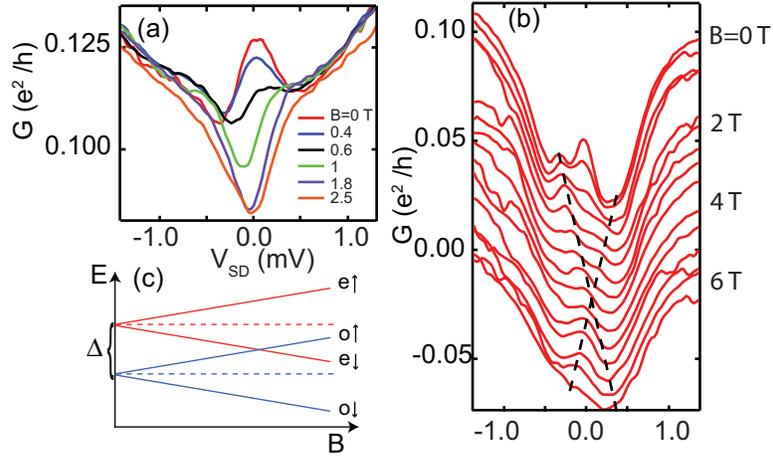}
\caption{\label{S2}(a) $B_\parallel$ dependence of the Kondo resonance in the first Coulomb diamond of Sample~II.  Here,  $V_T=-0.27$~V and $V_R=-0.05$~V, while $V_{g}=-0.02$~V is applied to gate M\@. (b) $B_\parallel$ dependence of the Kondo resonances in the second Coulomb diamond of Sample~II.    Here,  $V_T=-0.27$~V and $V_R=-0.05$~V as in (a), while $V_{g}=-0.04$~V is applied to gate M\@. The center peak is reduced by the magnetic field and the side peaks reveal information about both the valley and Zeeman splittings. The curves are offset for clarity.  (c) Schematic field-dependence energy diagram of the valley states in the QD.}
\end{figure}

Next we present measurements of Sample~II in parallel magnetic field $B_\parallel$ using an 8~$\mu$V ac bias. Gates T and R are kept at $V_T=-0.27$~V and $V_R=-0.05$~V respectively while gate M is swept by $V_g$. Again we find enhancement of conductance in two consecutive Coulomb diamonds. In one diamond the zero-bias conductance peak behaves in a similar fashion as Sample I, decreasing with increasing $B_\parallel$ field without splitting. As shown in Fig.~\ref{S2}(a) taken at $V_g=-0.02$~V, at $B_\parallel=0$ there is a single Kondo resonance. It gradually decreases as $B_\parallel$ increases and at $B_\parallel=2.5$~T the Coulomb blockade is fully recovered. \\

In an adjacent diamond with one fewer electron, some side peak features are resolved, suggesting comparable even and odd-valley coupling. In Fig.~\ref{S2}(b) taken at $V_g=-0.04$~V we show a series of conductance curves offset by (0.0015~$\frac{e^2}{h}/\text{T})\times \vert B_\parallel\vert$. The coexistence of the center resonance and the side resonance is consistent with a non-zero valley splitting. The center peak decreases with rising $B_\parallel$ without obvious splitting. There are traces of two side peaks, indicated by the dashed lines, which move towards the center. After converging at about 3.5~T they move away from each other again. These features are mainly contributed by the Zeeman splitting of the valley side peaks similar to that illustrated in Fig.~\ref{fig3}(b), revealing a zero-field valley splitting of about 0.34~meV. In Fig.~\ref{S2}(c) a schematic energy diagram of the valleys in the magnetic field is shown. The position of either side peak in Fig.~\ref{S2}(b) is determined by the relative distance between $o\uparrow$ and $e\downarrow$ in Fig.~\ref{S2}(c). The $g$ factors extracted from the two dashed lines (Fig.~\ref{S2}(b)) are about 1.8 and 1.5 respectively, smaller than but comparable to the standard value of $g=2$. The asymmetry inside the Coulomb blockade suppresses the Kondo peak at positive bias, resulting in a very weak signal that is hard to read. Nonetheless, the data are consistent with interplay between the spin and valley degrees of freedom, a phenomenon recently observed in ESR experiments in silicon quantum dots.\cite{Hao:2013}\\

\begin{figure}
\includegraphics{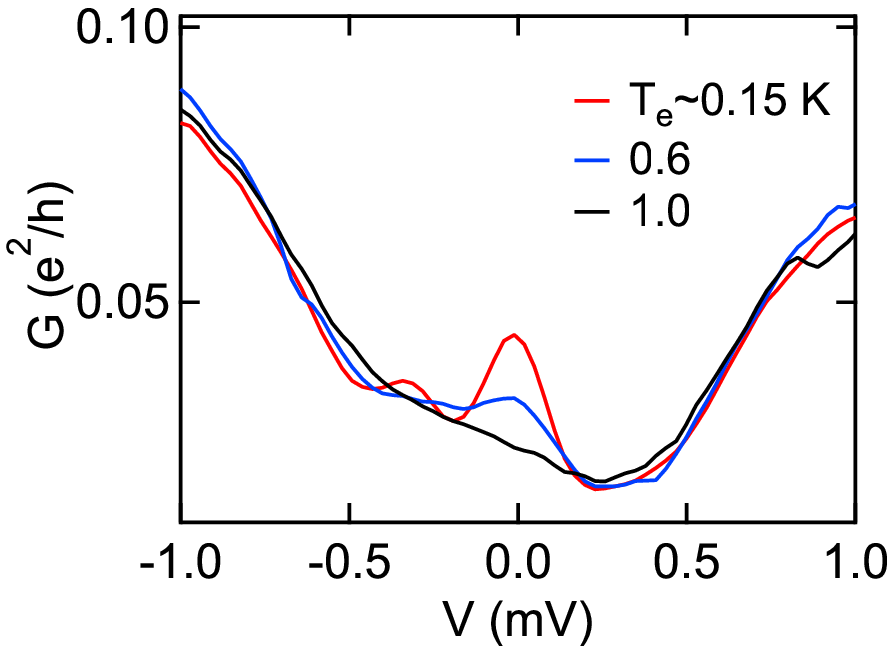}

 \caption{\label{S3}Temperature dependence of the Kondo resonances in Sample~II at zero magnetic field. Here,  $V_T=0.27$~V and $V_R=-0.05$~V, while $V_{g}=-0.04$~V is applied to gate M\@. }

\end{figure}

In Fig.~\ref{S3} we demonstrate at zero magnetic field the temperature dependence of the Kondo features shown in Fig.~\ref{S2}(b). Both the center peak and the side peak are reduced by rising temperature. At 0.6~K the peaks are significantly suppressed and the side peak becomes hard to discern. At 1.0~K both peaks are overcome by the Coulomb Blockade. The linewidth of the center peak, 0.18~mV, corresponds to a temperature of 2.1~K; for the side peak, the linewidth of 0.1~mV corresponds to 1.2~K.  These results are in reasonable agreement with the measured temperature dependence of the features.\\

In conclusion, we have measured Kondo effects in two separate Si/SiGe QDs. Kondo peaks that diminish with rising temperature are present in one pair of consecutive Coulomb diamonds in one sample (Sample~II), and in two separate pairs of consecutive diamonds in the other (Sample~I), for a total of six separate dot configurations. From side peak positions at zero field due to valley screening, zero-field valley splitting is revealed to be $\Delta\approx0.28$~meV in Sample~I and $\Delta\approx0.34$~meV in Sample~II. A zero-bias conductance peak that persists for finite magnetic field, which is consistent with the valley Kondo effect due to valley non-conservation in tunneling, is repeatedly observed in both samples, in a total of five separate Coulomb diamonds. The experiment suggests that the valley degree of freedom in Si can enrich the Kondo effect in Si/SiGe QDs and requires better theoretical understanding.\\












%






%





































\begin{acknowledgments}

We would like to thank J. Stettenheim for assistance with the dilution refrigerator. This work is supported at Dartmouth and UW-Madison by the Army Research Office under Grant No.\ W911NF-12-1-0607 and at Dartmouth by the NSF under Grant No.\ DMR-1104821. Development and maintenance of the UW-Madison growth facilities used for fabricating samples is supported by DOE Grant No.\ DE-FG02-03ER46028.

\end{acknowledgments}



%

\end{document}